# A nonparametric HMM for genetic imputation and coalescent inference

**Lloyd T. Elliott**[*] **and Yee Whye Teh**

*Department of Statistics*
*University of Oxford*
*24-29 St Giles' Rd.*
*Oxford, OX1 3LB, U.K.*
*e-mail:* elliott@stats.ox.ac.uk*;* teh@stats.ox.ac.uk

**Abstract:** Genetic sequence data are well described by hidden Markov models (HMMs) in which latent states correspond to clusters of similar mutation patterns. Theory from statistical genetics suggests that these HMMs are nonhomogeneous (their transition probabilities vary along the chromosome) and have large support for self transitions. We develop a new nonparametric model of genetic sequence data, based on the hierarchical Dirichlet process, which supports these self transitions and nonhomogeneity. Our model provides a parameterization of the genetic process that is more parsimonious than other more general nonparametric models which have previously been applied to population genetics. We provide truncation-free MCMC inference for our model using a new auxiliary sampling scheme for Bayesian nonparametric HMMs. In a series of experiments on male X chromosome data from the Thousand Genomes Project and also on data simulated from a population bottleneck we show the benefits of our model over the popular finite model fastPHASE, which can itself be seen as a parametric truncation of our model. We find that the number of HMM states found by our model is correlated with the time to the most recent common ancestor in population bottlenecks. This work demonstrates the flexibility of Bayesian nonparametrics applied to large and complex genetic data.

## 1. Introduction

Hidden Markov models (HMMs) are good approximations of the statistics of genetic sequences [1] and they have been applied extensively in genetic imputation and disease association studies (see [2] for a review). In these models, each genetic sequence in a population is assigned a succession of cluster indicators. At a given chromosome location, all genetic sequences with identical cluster indicators are genetically similar (*i.e.*, they originate from the same ancestral genetic material and thus they share many of the same mutations). These clusters can represent population components in an admixture or genetic founders in a population bottleneck. (Admixtures occur when isolated populations begin to interbreed and population bottlenecks occur when a population reduces in size for a time and then increases in size [3].)

---
[*]Corresponding author.

In this work, we present a Bayesian nonparametric HMM for genetic sequences based on the hierarchical Dirichlet process (HDP). Bayesian nonparametrics were originally developed to provide flexible and tractable priors on infinite dimensional probability spaces [4]. In this classical development, the Dirichlet process (DP) was used to associate a latent parameter with each observed data item. A draw from the DP posterior induces a clustering of the data items through the equivalence classes formed by the identity of the latent parameters: all data items with the same latent parameter are considered to be in the same cluster [5]. Bayesian nonparametric HMMs dynamically extend this idea by allowing the clustering of data items to vary along an axis such as time (or in the case of genetic data) location, and model the HMM states at each location using Dirichlet processes. To link cluster assignments between the locations of a Bayesian nonparametric HMM, the Dirichlet processes must be organized into a hierarchy in which the DPs at each location share a global set of clusters. This organization is an example of a hierarchical Dirichlet process [6]. The most general formulation of a Bayesian nonparametric HMM is known as the infinite HMM [7], or iHMM. The iHMM is specified by considering an infinite latent HMM state space, and placing a Dirichlet process prior on the transition kernel associated with each of the states.

The first applications of Bayesian nonparametric HMMs to genetic sequence data involved a direct application of the iHMM [8, 9]. In these works, the latent states of the iHMM correspond to haplotypes (haplotypes are collections of mutations that are inherited together by virtue of their proximity on the chromosome), and the iHMM was used to model arbitrary HMM transitions between haplotypes. This approach results in homogeneous HMMs with large parameter spaces.

Our model provides a novel formulation of a Bayesian nonparametric HMM as a non-homogeneous HMM in which the possible transitions are restricted to support biologically plausible parameterization. This yields a reduction in the size of the parameter space (relative to the iHMM), leading to more efficient inference and also improved imputation accuracy, and better biological interpretation of the latent parameters learned by the model. Our HMM is also designed to encourage HMM self transitions (as in the sticky HDP-HMM [10]), which capture the mosaic structure of haplotypes [11]. In addition, the non-homogeneous and restricted nature of our parameterization allows exact and truncation free Gibbs updates, whereas previous work in Bayesian nonparametric HMMs often relies on Metropolis-Hastings or truncated variational methods [10, 12].

The popular fastPHASE model [13] in statistical genetics can be seen as a finite truncation of our model, and so we will refer to our model as BNPPHASE (for Bayesian nonparametric fastPHASE). We develop a new sampling method for BNPPHASE based on a partially collapsed version of the HDP with auxiliary variables. This inference method is similar to beam sampling [12] for iHMMs, but with exact Gibbs updates rather than Metropolis-Hastings updates for the auxiliary variables.

In addition to iHMMs, nonparametric Bayesian fragmentation-coagulation processes (FCPs) have also been used to model genetic sequences [14, 15]. FCPs are Markov models in which a latent space of partitions transition along the chromosome through the splitting and merging of the blocks of the partitions. They can be seen as stationary Bayesian versions of the BEAGLE [16] model (which has a similar, although more heuristic, splitting and merging mechanic), and as such are well suited for producing haplotype graphs. However, BNPPHASE can model differential ancestry arising from admixture and bottlenecks by virtue of its nonhomogeneous HMM transitions (this nonhomogeneity is also a feature of fastPHASE [13]). The asymptotic complexity of inference for BNPPHASE is the same as that for FCPs and other nonparametric Bayesian methods based on HMMs: it is linear in the number of sequences and the number of chromosome locations, and linear in the expected number of HMM states, yielding an asymptotic complexity of $\mathcal{O}(NT \log NT)$, where $N$ is the number of individuals and $T$ is the number of chromosome locations considered.

We conducted two sets of experiments on genetic sequences using BNPPHASE. In the first set of experiments, we assessed the scalability of the BNPPHASE model and examined simulated data designed to model the out-of-Africa population bottleneck in humans. We found a strong negative correlation between the number of clusters found per site by both the BNPPHASE and fastPHASE models and the time to most recent common ancestor (TMRCA, or time to coalescence) in the simulated data. After regressing TMRCA against the number of clusters, residual error of the BNPPHASE model was smaller than that of fastPHASE. In the second set of experiments, we compared the accuracy of imputing missing information in male X chromosome from the Thousand Genomes Project [17] using BNPPHASE with that of fastPHASE, a Bayesian version of fastPHASE and BEAGLE. We considered two masking conditions for the imputation: a uniform condition, and a condition designed to mimic a whole genome study/reference paradigm in which individuals in a study panel are genotyped at a relatively small number of locations, and imputed into a reference panel of densely genotyped individuals. The BNPPHASE model performed competitively in the missing data imputation experiment in both masking conditions.

In the remainder of this section, we give a high level review of relevant aspects of statistical genetics, and provide a summary of the BNPPHASE model as well as a worked example. In Section 2, we provide details for the definition of our model as a Bayesian nonparametric HMM, and for our inference methods. In Sections 3 and 4 we describe our experiments and results and in Sections 5 and 6 we discuss our results and conclude.

### *1.1. The coalescent and recombination*

Under mild demographic assumptions (namely: no selection [18], random mating, constant population size, and uniform recombination rate [3]), the joint distribution of the mutations in a population is given by a process known as the coalescent with recombination [19]. Briefly, the coalescent with recombination is given as follows: the ancestry of $N$ genetic sequences in a population is formed by tracing their lineage backwards in time and placing coalescent events with rate $\frac{2}{N_e}\binom{k(\ell)!}{2}$, where $k(\ell)$ is the number of distinct lineages existing at time $\ell$, and $N_e$ is the effective population size [3]. At each coalescent event, two lineages chosen uniformly among all pairs of lineages are combined into one lineage. To account for shared ancestry arising from recombination events (in which a chromosome in an offspring is formed by recombining two parent chromosomes), recombination events are placed with rate $\rho k(\ell)/2$. Here, $\rho$ is a scaled recombination rate. At each recombination event, a lineage is chosen uniformly among all lineages and that lineage is split at a random point along the sequence to form two new ancestors for the lineage. All material to the left of the splitting point is inherited from one of the ancestors and all material to the right of the splitting point is inherited from the other ancestor. This process continues backwards in time until all the lineages have coalesced into the most recent common ancestor.

In the coalescent with recombination, the genealogy of a population (*i.e.*, the tree structure of the ancestry of a chromosome location) is location-dependent, and can be viewed as a genealogy-valued process indexed by chromosome location. Seen in this way, the transitions between genealogies are non-Markovian [20]. They are, however, modelled well by an approximating Markov assumption [1]. Failure to model the non-Markovian structure in the genealogy can lead to spurious associations in genetic association studies [21]. But the Markov assumption does not induce much loss of accuracy in inference [1] and allows linear time inference algorithms. This Markov assumption forms the basis of most HMM approaches for genetic sequence data (including parametric approaches, such as those reviewed in [2], and also nonparametric approaches).

An example of a simulation of 5 sequences from the coalescent with recombination viewed as a genealogy-valued process indexed by chromosome location is given in Figure 1 *(top)*. The genealogies presented in that figure are simulated from a bottleneck using the `ms` software [22]. The simulation parameters we used are from [23] and are designed to model the out-of-Africa bottleneck in humans. In that figure, due to the genealogies, sequences 3 and 4 would be genetically quite similar around location A, but quite different around locations B and C. In a reasonable HMM model of these sequences, we might expect the HMM state assignment of sequences 3 and 4 to be the same at location A, but to be different at locations B and C. For a deeper review of these aspects of statistical genetics, we refer to [3].

### *1.2. Mutations and alleles*

Most of the genetic material at a fixed chromosome location will be identical across all of the individuals in a population. This is due to the shared ancestry of the population. Differences in the material are only present at chromosome locations at which mutations have occurred more recently than the most recent common ancestor of the entire sample. Throughout this work, the mutations we will predominantly be interested in are single nucleotide polymorphisms (SNPs). A SNP is a chromosome location at which a mutation occurring in the ancestry of a population has lead to the observation of two possible DNA basepairs (called alleles) at that location (*i.e.*, it is a point mutation). The more common allele is called the major allele and the less common allele is called the minor allele. The SNP data for a population can therefore be summarized by the matrix $x = ((x_{it})_{t=1}^{T})_{i=1}^{N}$ for which $x_{it} = 1$ indicates that sequence $i$ has the minor allele at location $t$ and $x_{it} = 0$ indicates that sequence $i$ has the major allele at location $t$. We also note that under the genetic assumptions listed in the previous subsection, the distribution of unique mutation patterns in a population is given locally by a DP (see [24] for more detail). This observation further recommends the use of nonparametric Bayesian priors based on the Dirichlet process for genetic models.

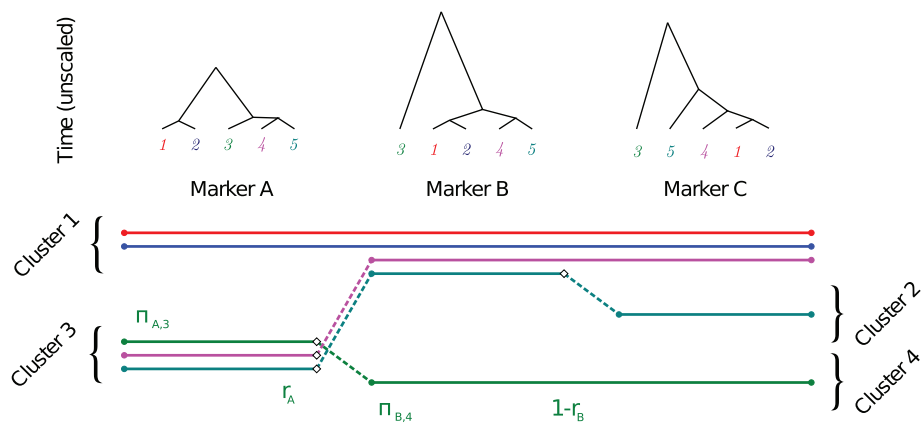

FIG 1. Top: *Genealogy of 5 genetic sequences typed at locations A, B and C with simulated ancestry. y-axis indicates time to coalescence. x-axis indicates location. Coalescence of lineage indicates common ancestry.* Bottom: *Illustration of BNPPHASE model. Color indicates sequence identity (with red being sequence 1, blue being sequence 2, ...). Dotted lines indicate cluster transitions. y-axis indicates cluster assignments: sequences that are in the same cluster at a location have adjacent y-levels.*

Humans are diploid organisms: most of our chromosomes come in pairs. Presently, low cost DNA sequencing methods are unable to determine, out of a matched pair of chromosomes, which of the two chromosome a given allele originates from. Instead, these methods report only the presence or absence of an allele among either of the matched chromosomes. The process of assigning alleles to one of two matched chromosomes is known as phasing. Genetic sequence data which has been phased is known as phased data. In this work, we will deal only with phased data, although the possible extension of our method to phasing is clear: the required change is an extension of the message pass-ing we use for inference to pairs of chromosomes as is done in [13]. Sources of ground truth for phased data in humans include male X chromosome data and simulated data.

### 1.3. The BNPPHASE model

Suppose that alleles for $N$ genetic sequences from a population are observed at $T$ SNPs. The BNPPHASE model associates a latent cluster assignment to each sequence and location (we will denote the cluster assignment of the $i$-th individual at the $t$-th location by $z_{it}$, which will be supported on all natural numbers). Between each pair of consecutive locations, a given sequence either remains in the same cluster, or with some probability proportional to a rate $r_t$ (which depends on chromosome location) is reassigned to one of the clusters with a location-dependent probability. The tendency for the cluster assignment to persist between sequence coordinates (as in sticky HMMs [10]) reflects local smoothness of the genealogy.

The rate $r_t$ governs the dependence among the clusterings: if $r_t = 0$ then the clusterings at $t-1$ and $t$ are same and if $r_t = 1$ then the clusterings are at $t-1$ and $t$ are independent. Intuitively, we want to infer small values of $r_t$ on regions of the sequence for which the underlying genealogy structure from the coalescent does not change much, and larger values of $r_t$ for locations where recombination events in the ancestry of the population have lead to substantial changes in the latent genealogy structure. (For the recombination hotspots described in [25] $r_t$ should be larger.)

We introduce latent variables $y_{it}$ indicating if the $i$-th sequence has a transition event after the $t$-th location. We also introduce cluster weights $\pi_{tk}$ for the $k$-th cluster at location $t$ (such that $\pi_{tk} \geq 0$ and $\sum_k \pi_{tk} = 1$). The cluster assignment of the $i$-th sequence at position $t$ is *a priori* drawn from a discrete distribution with the probability of $z_{it} = k$ being $\pi_{tk}$ in the case where $y_{i,t-1} = 1$. Otherwise (if $y_{i,t-1} = 0$) the cluster assignment of the $i$-th sequence at position $t$ is copied from $z_{i,t-1}$. For the first location, the prior distribution on the cluster assignment of the $i$-th individual is given by $\pi_{1k}$. The location-dependence of $\pi$ lets BNPPHASE capture changes in the genealogy along the chromosome induced by the coalescent with recombination. If the transition probabilities of an HMM depend on location (as is the case for our $\pi$), then the HMM is called a non-homogeneous HMM. In contrast, other nonparametric models for genetic variation do not involve location dependent priors, and are homogeneous HMMs.

The number of clusters at each location, and the prior distribution over the local cluster weights $\pi_{tk}$, are given by a Dirichlet process (DP). In order to make this distribution well defined, we have to identify the clusters at each location $t$ with global clusters that persist across the whole process. This is done in Section 2 by introducing an HDP. The HDP is the simplest method for linking together cluster proportions such that the number of clusters is unbounded and the induced prior distribution on the cluster assignments does not depend on the order in which the study individuals are observed or the size of the population from which the study individuals were selected (*i.e.*, it is exchangeable and projective) [6]. An illustration of the BNPPHASE model is provided in Figure 1 (*bottom*).

Given a fixed setting of the latent variables and parameters of BNPPHASE, we will model the alleles ($x_{it}$) as a matrix of independent Bernoulli random variables. The distribution of each entry $x_{it}$ depends only on the cluster assignment ($z_{it}$) of the $i$-th sequence at location $t$, and the BNPPHASE parameters. In particular, if the $i$-th sequence is assigned to cluster $k$ at location $t$, then $x_{it} \sim \text{Bernoulli}(\theta_{tk})$. Here, $\theta_{tk} \in [0,1]$ is a parameter associated with the $k$-th cluster and the $t$-th location.

We place a hierarchical prior on $\theta$ as follows: $\theta_{tk}$ is drawn from a beta distribution with mean and mass which both depend on $t$, so $\theta_{tk} \sim \text{Beta}(\gamma_t \beta_t, \gamma_t(1-\beta_t))$. The mean $\beta_t$ is drawn from the beta distribution $\text{Beta}(b,b)$ in which $b$ is a global parameter controlling the variance of the allele frequencies. We place exponential priors with rate 1 on $b$ and on each of the local masses $\gamma_t$. The motivation for this hierarchical likelihood is two fold. First, marginalizing $b, \gamma$ and $\beta$ produces a distribution on $\theta$ with heavier tails than would be produced by a beta distribution with fixed mean and variance, allowing more flexibility for modelling rare variants, which has been shown to be important in haplotype models [26]. Second, sharing the $b$ parameter across multiple locations allows the model to learn a global allele frequency parameter. This likelihood can easily be generalized to multi-allelic mutations such as copy number variations or micro-satellites by replacing the beta/Bernoulli model with a discrete/Dirichlet (these more complicated types of mutation are described in [3]).

*Illustration of BNPPHASE*

In Figure 1 (*bottom*), we suppose that our data consists of five phased genetic sequences typed at three biallelic locations (labelled as A, B and C). The sequences are labelled by color: red, blue, green, magenta and teal. The latent cluster assignment for BNPPHASE has represented the data using four latent clusters corresponding to the $y$-level of the sequences in Figure 1 (*bottom*). At the first location, BNPPHASE has clustered the data into two clusters: with the red and blue sequences in cluster 1 (contributing $\pi_{A,1}^2$ to the probability), and remaining sequences in cluster 3 (contributing $\pi_{A,3}^3$ to the probability). Between location A and location B, the green, magenta and teal sequences transit to new clusters contributing $(1-r_A)^2 r_A^3$ to the probability (the fact that the red and blue sequences do not transit contributes an additional $(1-r_A)^2$). Between location A and B, the green, magenta and teal sequences have transitioned to clusters 1, 1 and 4 respectively contributing $\pi_{B,1}^2 \pi_{B,4}$ respectively to the prior. The probabilities for the transitioning and clustering between locations B and C can similarly be read off of Figure 1 (*bottom*) resulting in a total probability (conditioned on $\pi$ and $r$) of the illustration in Figure 1 (*bottom*): $(1-r_A)^2 r_A^3 (1-r_B)^4 r_B \pi_{A,1}^2 \pi_{A,3}^3 \pi_{B,1}^2 \pi_{B,4} \pi_{C,3}$. The prior on $\pi$ is given by a hierarchical Dirichlet process, which will be presented formally in subsection 2.1.

*1.3.1. Relationship to fastPHASE*

If the BNPPHASE model were truncated such that only $K$ clusters were used for a fixed finite $K$, then the prior induced by BNPPHASE on the clustering of the genetic sequences would reduce to the following finite form:

$$\Pr(y, z, \pi) = \Pr(\pi) \prod_i \pi_{1,z_{i1}} \prod_{i,t>1} \begin{cases} r_{t-1} \pi_{t,z_{it}} & y_{i,t-1} = 1, \\ 1 - r_{t-1} & y_{i,t-1} = 0, z_{it} = z_{i,t-1}, \\ 0 & \text{otherwise.} \end{cases} \quad (1)$$

This is a Bayesian version of fastPHASE [13] with a prior given by $\Pr(\pi)$. In [13], the parameter $\pi$ for equation (1) is estimated using an ML fit. In [27], a Bayesian version of equation (1) is proposed with a Dirichlet distribution prior on $\pi$ with a fixed number of components, but with a discrete distribution on the allele emissions (*i.e.*, $\theta$ was constrained to be 0 or 1). Thus, we can view the BNPPHASE model as a nonparametric Bayesian extension of [27] and [13].

## 2. Methods

In this section, we give some theoretical background on Bayesian nonparametrics and then we describe the BNPPHASE model as a generative nonparametric process. Finally, we describe an MCMC inference algorithm for BNPPHASE.

### 2.1. The hierarchical Dirichlet process

The HDP can be used to define a prior distribution on cluster proportions. In the BNPPHASE model, the HDP is used to provide cluster proportions at each chromosome location. The joint distribution over the cluster proportions induced by this prior is given by the following hierarchical stick breaking process from [6]:

$$H = \sum_{k=1}^{\infty} \omega_k \delta_{\chi_k}, \qquad \omega_k = \eta_k \prod_{k'=1}^{k-1}(1-\eta_{k'}), \qquad \eta_k \stackrel{i.i.d.}{\sim} \text{Beta}(1, \alpha_0), \quad (2)$$

$$G_t = \sum_{k=1}^{\infty} v_{tk} \delta_{\psi_{tk}}, \qquad v_{tk} = \xi_{tk} \prod_{k'=1}^{k-1}(1-\xi_{tk'}), \qquad \xi_{tk} \stackrel{i.i.d.}{\sim} \text{Beta}(1, \alpha), \quad (3)$$

$$\chi_k \stackrel{i.i.d.}{\sim} \text{Uniform}[0,1], \qquad \psi_{tk} \stackrel{i.i.d.}{\sim} H, \qquad \pi_{tk} = \sum_{k': \psi_{tk'} = \chi_k} v_{tk'}. \quad (4)$$

Here $\alpha > 0$ and $\alpha_0 > 0$ are concentration parameters, $1 \leq t \leq T$ index the chromosome locations and $\delta_x$ denotes an atomic mass at $x$, and $H$ and $G_t$ are Dirichlet processes. The distribution on the vector $\omega = (\omega_1, \omega_2, \ldots)$ is known as the GEM distribution [29], and the distribution on $\pi_t = (\pi_{t1}, \pi_{t2}, \ldots)$ is a coagulation of the GEM distribution [29, 30].

In models based on the Dirichlet process, the support of the atomic mass locations $\chi_k$ are usually taken to be the parameter space of the model. For our situation, this would be the space of allele emission probabilities $\theta_{k1}, \ldots, \theta_{kT}$. As it is easier to describe our model using the GEM view of the HDP, we will instead link the atomic masses to the allele emission probabilities through their indices $k$, and discard $\chi_k$.

## 2.2. Generative process for BNPPHASE

The BNPPHASE model can be described by a generative process conditioned on the hyperparameters $\alpha_0$, $\alpha$, $b$ and $\gamma_t$:

1. Draw $\omega|\alpha_0$ from the GEM distribution (equation 2).
2. For $1 \leq t \leq T$: draw $\pi_t|\omega, \alpha$ from the coagulation of the GEM (given by equations 3 and 4).
3. For $1 \leq t < T$: draw $r_t \sim \text{LogUniform}(r_{\min}, 1)$.
4. For $1 \leq i \leq N$:
    (a) Draw $z_{i1}$ from $\pi_1$.
    (b) For $1 \leq t < T$: draw $y_{it} \sim \text{Bernoulli}(r_t)$ and:
        i. If $y_{it} = 1$: draw $z_{i,t+1}$ from $\pi_{t+1}$.
        ii. Else if $y_{it} = 0$: set $z_{i,t+1}$ to $z_{it}$.
5. For $1 \leq t \leq T$: draw $\beta_t \sim \text{Beta}(b, b)$.
6. For $1 \leq t \leq T, k \geq 1$: draw $\theta_{tk} \sim \text{Beta}(\gamma_t \beta, \gamma_t(1-\beta))$.
7. For $1 \leq t \leq T, 1 \leq i \leq N$: draw $x_{it} \sim \text{Bernoulli}(\theta_{t,z_{it}})$.

Here, $r_t \sim \text{LogUniform}(a, b)$ means that $r_t$ is a random variable such that $\log r_t$ is uniformly distributed on the interval $[\log a, \log b]$. We chose this weakly informative heavy tailed prior on $r_t$ so that a single cluster can extend over large chromosome regions (over which $r_t$ has small values) while still allowing recombination hotspots [25] to occur (these are locations for which $r_t$ is relatively large). We place a log Normal prior on $\alpha_0$ and $\alpha$:

$$\alpha_0 \sim \text{LogNormal}(\log \alpha_{0\text{mean}}, \alpha_{0\text{var}}), \tag{5}$$
$$\alpha \sim \text{LogNormal}(\log \alpha_{\text{mean}}, \alpha_{\text{var}}). \tag{6}$$

Here $\alpha_{0\text{mean}}, \alpha_{0\text{var}}, \alpha_{\text{mean}}$ and $\alpha_{\text{var}}$ are all fixed positive parameters that control the number of nonempty clusters in the BNPPHASE prior. The priors on $\gamma, \beta$ and $b$ were described above in Section 1.3, and so this concludes the specification of the BNPPHASE model. A graphical model depicting the conditional dependence relationships among the latent variables of the BNPPHASE model is provided in Figure 2. This figure shows the HMM structure governing dependence between adjacent location dependent variables, and also the hierarchical organization of the Dirichlet processes.

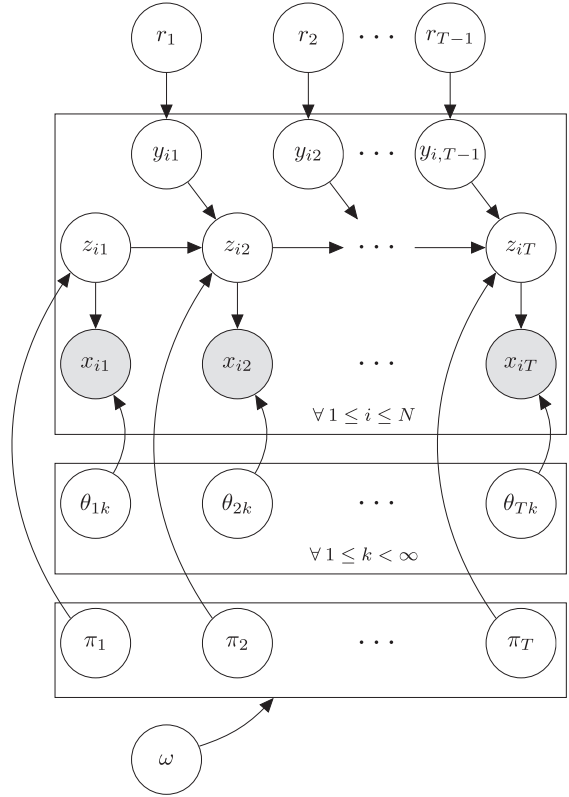

FIG 2. *Graphical model for BNPPHASE. White circles indicate latent variables (learned by BNPPHASE). Grey circles indicate observed alleles. Arrows indicate conditional dependence relationships (described as in plate diagram notation from [28]). Parameters $\alpha, \gamma, \beta$ and $b$ suppressed for brevity.*

## *2.3. Inference for the BNPPHASE model*

We use MCMC to conduct posterior sampling of the BNPPHASE model conditioned on observed data. In the experiments that follow, we will be interested in the conditional distribution of missing alleles, and in the posterior distribution over the number of clusters found by the BNPPHASE model. These statistics are estimated by averaging over all MCMC samples produced after a number of burnin iterations have been completed.

For a fixed sequence $i$, we update the cluster assignments $z_{it}$ and transitions $y_{it}$ for $t = 1, \ldots, T$ using the forwards filtering/backwards sampling algorithm [31] along with a bespoke auxiliary Gibbs update designed to efficiently handle the infinite state space of the HDP. The parameters $\alpha_0, \alpha, r_t, b, \gamma_t$ and $\beta_t$ are updated using slice sampling [32]. The likelihood parameters $\theta_t$ are integrated out. In the remainder of this section, we derive the auxiliary Gibbs update for the cluster assignments and transitions. Unlike previous samplers for HMMs based on the HDP [12, 10], the nonhomogeneous nature of the BNPPHASE model allows the derivation of truncation free and exact updates.

## 2.4. Auxiliary Gibbs update for $z_{it}, y_{it}$

We derive an auxiliary Gibbs update for $z_{it}, y_{it}$ based on a representation of the law in equations (2, 3 and 4) which uses the CRP distribution, a distribution on partitions. If $S$ is a set then a partition $\mathcal{R}$ of $S$ is a set of nonempty subsets of $S$ (called blocks) whose union is $S$. A random partition $\mathcal{R}$ of $S$ is distributed as a CRP with concentration $\alpha$, denoted $\mathcal{R} \sim \mathrm{CRP}(S, \alpha)$, if the probability distribution function for $\mathcal{R}$ is given as follows:

$$\Pr(\mathcal{R}) = \alpha^{|\mathcal{R}|} \Gamma(\alpha)/\Gamma(\alpha+|S|) \prod_{a \in \mathcal{R}} \Gamma(|a|). \qquad (7)$$

Here $|X|$ denotes the size of the set $X$. The CRP can also be expressed as a sequential scheme in which the partition $\mathcal{R}$ is built by adding one element at a time such that the first item forms a block by itself and item $i$ joins a block $a$ with probability $|a|/(i+\alpha-1)$ or forms a block by itself with probability $\alpha/(i+\alpha-1)$ (see [6] or [33] for more details).

Suppose that we draw $\omega, (\pi_t)_{t=1}^T | \alpha, \alpha_0$ according to the coagulated GEM distribution (equations 2, 3 and 4) and we draw $z_{it}$ according to the law $\Pr(z_{it} = k) = \pi_{tk}$ for $i \in S_t$, $1 \leq t \leq T$. This procedure is equivalent to the following scheme: draw $\omega_k | \alpha_0$ according to (2), then draw $\mathcal{R}_t \sim \mathrm{CRP}(S_t, \alpha)$, and finally for each block $\zeta \in \mathcal{R}_t$ draw $\varphi_{t\zeta}$ according to the probability $\Pr(\varphi_{t\zeta} = k) = \omega_k$, and set $z_{it} = \varphi_{t,\zeta_{it}}$ where $\zeta_{it}$ is the block of $\mathcal{R}_t$ containing $i$. (Note that the first index of $\varphi$ is an integer denoting the location at which the partition is placed, and the second index is a set—a block of a partition.) The equivalence of these two schemes is proven in [6] and [30], and allows us to expresses the HDP using a DP at the top level and CRPs at the bottom level.

Due to this equivalence, we can use CRPs to derive a new generative process for the cluster assignments of the BNPPHASE model in which the individuals are sampled sequentially, conditioned on $\omega$. We do this by describing a sequential scheme for generating the cluster assignments and jumps of the $i$-th sequence, conditioned on all of the following variables: $\omega$, the cluster assignments and jumps of all sequences $i' < i$, and also the auxiliary variables: the partitions $\mathcal{R}_t$ restricted to the first $i-1$ sequences (we denote this restriction by $\mathcal{R}_t^{-i}$), and the assignment of the blocks of the partitions $\mathcal{R}_t^{-i}$ to coordinates of the GEM distribution $\omega$ (i.e., the $\zeta_{i't}$ for $i' \neq i$). We will refer to all of these conditioning variables as 'rest'. A graphical model illustrating the conditional dependences for the BNPPHASE model among the augmented set of variables ($\zeta, \varphi$ and $S$) in this extended representation is provided in Figure 3.

When $y_{i,t-1} = 1$, $z_{it}$ must be sampled from $\pi_t$. Under the CRP representation, this can be achieved by adding individual $i$ to the partition $\mathcal{R}_t^{-i}$ (which has distribution $\mathrm{CRP}(\{i' < i : y_{it} = 1\}, \alpha))$. If individual $i$ joins a block $\zeta$ that already exists in $\mathcal{R}_t^{-i}$ (we denote this event by $\zeta_{it} = \zeta$), then we set $z_{it} = \varphi_{t\zeta}$. If the $i$-th individual forms a new block in the partition $\mathcal{R}_t^{-i}$ (we denote this event by $\zeta_{it} = \varnothing$), then we must sample $\varphi_{t\varnothing}$, and set $z_{it} = \varphi_{t\varnothing}$. When $y_{i,t-1} = 0$, $z_{it}$ is set to $z_{i,t-1}$ and $\zeta_{it}$ is set to 0. (The $\zeta_{it} = 0$ notation means that $\zeta_{it}$ was not actually sampled from $\pi_{it}$ for individual $i$ at location $t$ because that individual did not transition immediately before location $t$). This induces the following conditional distribution:

$$\Pr((z_{it}, \zeta_{it})_{t=1}^T | \text{'rest'}) = J_1(z_{i1}, \zeta_{i1}) \prod_{t=2}^T \begin{cases} r_{t-1} J_t(z_{it}, \zeta_{it}) & \text{if } \zeta_{it} \neq 0, \varphi_{t\zeta_{it}} = z_{it}, \\ 1 - r_{t-1} & \text{if } \zeta_{it} = 0, z_{it} = z_{i,t-1}, \\ 0 & \text{otherwise.} \end{cases}$$

$$\cdot \prod_{t=1}^T L_{it}(x_{it} | z_{it}, \text{'rest'}) \qquad (8)$$

$$\text{Where } J_t(z_{it}, \zeta_{it}) = \frac{1}{|R_t^{-i}| + \alpha} \cdot \begin{cases} |\zeta| & \zeta_{it} = \zeta \in \mathcal{R}_t^{-i}, \\ \alpha \omega_z, & \zeta_{it} = \varnothing, z_{it} = z, z \in \mathbb{N}, \\ 0 & \text{otherwise.} \end{cases} \quad (9)$$

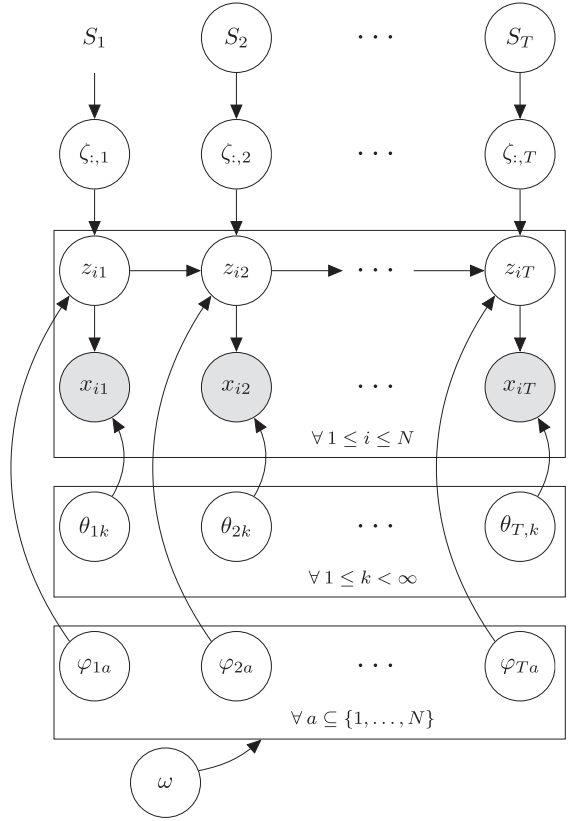

FIG 3. *Graphical model for augmented version of BNPPHASE including variable $\zeta$ indicating local block assignments of individuals. Sets $S_t$ indicate which individual assignments 'jump' and are reassigned before the t-th location. $S_1 = \{1, \ldots, N\}$, as all individuals are initially assigned at location $t = 1$. Model parameters suppressed for brevity.*

Here $J_t(z, \zeta)$ is the marginal probability that sequence $i$ 'jumps' to $z$ at location $t$, and $L_{it}(x_{it}|z_{it},\text{'rest'})$ is a likelihood formed by marginalizing $\theta$:

$$L(x_{it}|z_{it} = k) = \frac{1}{\gamma_t + n_{1tk}^{-i} + n_{0tk}^{-i}} \cdot \begin{cases} \gamma_t \beta_t + n_{1tk}^{-i} & \text{if } x_{it} = 1, \\ \gamma_t(1 - \beta_t) + n_{0tk}^{-i} & \text{if } x_{it} = 0. \end{cases} \quad (10)$$

Here $n_{1tk} = |\{i : z_{it} = k, x_{it} = 1\}|$ and $n_{0tk} = |\{i : z_{it} = k, x_{it} = 0\}|$ denote the counts of the number of times each allele is observed among the sequences assigned to each cluster.

Using exchangeability, we derive a forwards filtering/backwards sampling [31] message passing algorithm to sample the posterior of the cluster assignment of the $i$-th sequence using the conditional GEM probabilities multiplied by the hierarchical allele likelihood described in the generative process in Section 2.2. This algorithm is provided in the appendix.

## 2.5. Slice sampling for parameters $\alpha_0, \alpha, \gamma_t, \beta_t, b$ and $r_t$

Slice sampling provides efficient updates for random variables for which probability distributions are known only up to a normalizing constant, without requiring a choice of a proposal distribution or step size. We will update the variables $\alpha, \alpha_0, \gamma_t, \beta_t, b$ and $r_t$ using slice sampling. The unnormalized probability density functions of these variables are given in this section (by Bayes' rule, these unnormalized density functions are proportional to their conditional distributions). For more detail on slice sampling, we refer to [32].

*Conditional distributions for $\alpha_0$ and $\alpha$*

These distributions follow from the generative process in Section 2.2 and the CRP marginals for $\mathcal{R}_t$ and the definition of the GEM distribution [34].

$$\Pr(\alpha_0|\mathcal{R}, \varphi, K) \propto \Pr(\alpha_0) \alpha_0^K \Gamma(\alpha_0) \Big/ \Gamma\left(\alpha_0 + \sum_{t,k} |\{a \in \mathcal{R}_t : \varphi_{ta} = k\}|\right), \quad (11)$$

$$\Pr(\alpha|\mathcal{R}) \propto \Pr(\alpha) \prod_{t=1}^{T} \frac{\alpha^{|\mathcal{R}_t|} \Gamma(\alpha)}{\Gamma(\alpha + |R_t|)}. \quad (12)$$

Here $K$ is the number of unique non-empty clusters in the top level of the Dirichlet process hierarchy (i.e., $K = |\cup_{t=1}^{T} \{\varphi_{ta} : a \in \mathcal{R}_t\}|$).

*Conditional distributions for $\gamma_t, \beta_t, b$ and $r_t$*

By Bayes' rule, the unnormalized conditional distributions for $\gamma_t, \beta_t$ and $b$ are clear from their definition as a hierarchical likelihood, allowing slice sampling to be specified. Finally, a slice sampling update for $r_t$ for $1 \leq t < T$ is provided by the following conditional distribution:

$$\Pr(r_t|R_t) \propto \Pr(r_t) r_t^{|R_{t+1}|} (1 - r_t)^{N - |R_{t+1}|}. \tag{13}$$

Here $\Pr(r_t)$ is the prior on $r_t$ from the generative process in Section 2.2. This concludes the specification of slice sampling for $\alpha_0, \alpha, \gamma_t, \beta_t, b$ and $r_t$. Combined with the algorithm for updating $z_{it}, y_{it}$ in the appendix, we have therefore provided an MCMC algorithm for resampling the parameters and the cluster assignments from their posterior distribution under the BNPPHASE model, conditioned on the allele matrix $((x_{it})_{t=1}^T)_{i=1}^N$. Software implementing this MCMC scheme is available at https://github.com/BigBayes/BNPPhase (this software is provided under the BSD 2-clause license).

The asymptotic complexity of a single MCMC iteration is $\mathcal{O}(NTK)$. This can be computed by examining the equations for the message passing and the slice sampling. This complexity depends on $K$, which is random. In the prior, $K = \mathcal{O}(\log NT)$, but in the posterior the asymptote for $K$ depends on the data.

## 3. Experiments

We conducted two sets of experiments in which we compared the BNPPHASE, fastPHASE and various other baselines. In our first set of experiments, we examined the BNPPHASE posterior and the runtime of our MCMC inference using simulated data. In experiment *Ia)*, we used parameters from [23] to simulated data designed to model the out-of-Africa bottleneck in humans. We simulated 500 phased genetic sequences on 20 independent chromosome regions. Each region was on average $3 \times 10^5$ base pairs long. There were on average 2,100 biallelic locations in each region. We recorded the TMRCA of each biallelic location under the simulation and conducted inference of the latent clustering structure of the fully observed bottleneck data using fastPHASE and BNPPHASE. We then regressed the TMRCA against the number of clusters each model used per location. The number of clusters used by the fastPHASE model was computed by counting non-empty clusters in the maximum likelihood cluster assignment found by the EM algorithm for fastPHASE [13].

In experiment *Ib)*, we recorded the runtime our posterior MCMC inference conditioned on data simulated from the prior of the BNPPHASE model. The parameters used for simulating these data were as follows: $\alpha_{0\text{mean}} = 10.0$, $\alpha_{0\text{var}} = 1.0$, $\alpha_{\text{mean}} = 1.0$, $\alpha_{\text{var}} = 1.0$ and $r_{\min} = 10^{-5}$. In this runtime experiment, we varied the number of individuals between 100 and 900 and varied the number of sites between 100 and 900. For the trials in which the number of individuals were varied, we fixed the number of sites at 200 (and *visa versa* for the trials in which the number of sites were varied). For each combination of sites and individuals, we conducted 10 trials of 200 MCMC iterations each, and recorded the runtime of each trial.

In our second set of experiments, we performed imputation on male X chromosome data, and compared the imputation accuracies of BNPPHASE, fastPHASE and BEAGLE. In experiment *IIa)*, we formed a collection of datasets consisting of 20 intervals chosen randomly from the non-pseudoautosomal region of the male X chromosome. Each of these datasets contained 500 consecutive SNPs (an average length of around $10^5$ basepairs) from 524 male X chromosomes from the Thousand Genomes Project [17]. We held out nested sets of between 10% and 90% entries uniformly and examined imputation accuracy using fastPHASE, BEAGLE and BNPPHASE. Due to limitations in the fastPHASE software, only an even number of male X chromosomes can be considered by the fastPHASE software, and so in each of the 20 datasets we randomly discarded one of the 525 male X chromosomes contained in the Thousand Genomes Project.

In experiment *IIb)*, we performed a whole chromosome analysis of the non-pseudo-autosomal region of all 525 male X chromosomes from the Thousand Genomes Project. We partitioned the entire pseudo-autosomal region into 1,251 chunks each with at most 1,000 contiguous SNPs (the first 1,250 chunks contained 1,000 SNPs each and the last chunk contained 218 SNPs). We chose 262 study individuals at random, and in each chunk we held out 50% of the SNPs in the chunk for those study individuals. This hold-out condition mimics the study/reference paradigm used to impute a relatively sparsely assayed study sample into a reference panel [35]. We compared the imputation accuracy of BNPPHASE with that of a variational Bayes version of fastPHASE (VBFP) based on [13] and [27], with a factorized approximation of the fastPHASE distribution given in equation (1). For VBFP, we placed a finite Dirichlet distribution prior with $K$ components on $\pi$ (as in [27], so $\pi_t \sim \beta(a_1, \ldots, a_K)$). Additionally, we placed a Beta prior on $\theta$ (so, $\theta_{tk} \sim \text{Beta}(u_{tk1}, u_{tk2})$). We treated as parameters the jump rate $r$, the parameters of the Dirichlet distribution $(a_1, \ldots, a_K)$ and the parameters of the Beta distribution in the heirarchical specification of the parameters on $\theta$ ($u_{tk1}$ and $u_{tk2}$) and update them with M-steps. The latent variables $\pi$, $z$ and $\theta$ were represented with approximating distributions and updated using E-steps. The E-steps are identical to those derived in [13] (although restricted to the case of phased data), and the M-steps were found with standard methods [36]. The VBFP model was designed to be closer to the BNPPHASE model than the original fastPHASE specification in [13], and so we expect differences in the performance of VBFP and BNPPHASE to indicate more closely the advantage of adding a Dirichlet process prior.

### 3.1. MCMC initialization, burnin, iteration, restarts and parameter update schedule

The procedure we used for simulating the posterior of the BNPPHASE model with MCMC in experiments *Ia)* and *Ib)* was as follows. First, we initialized the chain using a scheme in which one sequence of the chain was initialized at a time, conditioned on previously initialized sequences. This initialization method is similar to the product of approximate conditionals method in [37]. Subsequently, 50 MCMC iterations were performed consisting of full sweeps over the parameters and Gibbs updates for the latent cluster assignments and jump indicators of each sequence. In these iterations, the parameters were updated 10 times for each single update of the latent state assignments and jump indicators. The first 20 of these iterations were discarded as burnin. This procedure was repeated 25 times, each time with an independent initialization, yielding 750 iterations which were averaged to produce posterior predictions.

TABLE 1
*RMSE for regression of TMRCA against number of clusters for BNPPHASE model, fastPHASE with default parameters (FP) or $K = 200$ clusters (FP200), and minor allele frequency (MAF).*

| Method | BNPPHASE | FP | FP200 | MAF |
|---|---|---|---|---|
| RMSE | **0.2724** | 0.2855 | 0.3063 | 0.3215 |

We chose the number of iterations to use by looking at trace plots of the likelihood and accuracy. These traces plateaued at around 50 iterations, after which a reasonable mode was found. Other methods in genetic imputation use similar numbers of iterations. Default parameters for fastPHASE and BEAGLE are 25 and 10 iterations respectively (including any burnin). The small number of iterations required for HMM methods in genetic imputation suggest that for haplotype models the posterior is quite peaked over its mode.

For experiment *IIb)*, we used the same MCMC schedule as described above for BNPPHASE, but with just one random restart (so, a total of 30 MCMC iterations collected per chromosome chunk). The VBFP was run for 60 iterations, of which the posterior predictions of the last 30 iterations were averaged. In every case, accuracy was computed by thresholding posterior predictions and recording the proportion of correctly called alleles. All of these experiments were performed on an AMD 1.9GHz Opteron processor, with one cor per random restart.

## 4. Results

### 4.1. Experiment Ia) simulated bottleneck

We found a strong negative correlation between the number of clusters used per location and the TMRCA for both the BNPPHASE model and fastPHASE. In Figure 5 *(left, right)* we regress the TMRCA against the number of clusters used per location. When we ran fastPHASE with default settings, fastPHASE would almost always choose to use 20 clusters (the maximum) in the ML cluster assignment. When we increased the maximum number of clusters to 200 (but otherwise left the parameters of fastPHASE with their default settings) large numbers of clusters were still used (as can be seen in Figure 5). BNPPHASE often used fewer clusters than fastPHASE. Visual inspection of these data suggests that fewer clusters (on the order of the numbers used by BNPPHASE) are often more reasonable representations of the data, as in Figure 8 (*top*). As a control, we regressed the TMRCA against the minor allele frequency and in this case we also found a negative correlation. The residual root mean squared errors of the regression were smallest in the BNPPHASE model (Table 1).

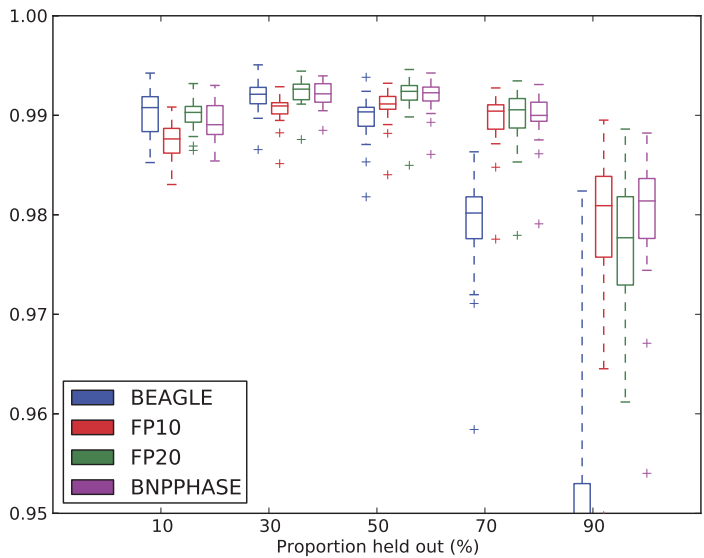

FIG 4. *Imputation accuracy for X chromosomes from the Thousand Genomes Project [17], uniform hold out condition. y-axis is truncated to emphasize variation over the 10%-70% hold out range.*

### 4.2. Experiment Ib) runtime on simulated data

In Figure 6 we show the runtime of the BNPPHASE model on simulated data drawn from the BNPPHASE prior. A linear dependence of runtime on both the number of individuals (Figure 6 *left*) and the number of sites (Figure 6 *right*) is clear from this figure. This result is consistent with our calculation of the asymptotic complexity of the BNPPHASE model, in which the dominant term was found to be $NT$.

### 4.3. Experiment IIa) male X chromosomes, uniform hold out

In Figure 8 we show an example region of the male X chromosome used in the imputation experiment on data from the Thousand Genomes Project. Figure 8 *(top left)* shows the pattern of minor alleles in this example region. In Figure 8 *(top right)*, a single sample from an MCMC chain for the BNPPHASE posterior is displayed. By comparing this sample with Figure 8 *(top left)*, we note that unique patterns of alleles in the data locally correspond to clusters in the BNPPHASE posterior: it is therefore clear that the clustering structure found by BNPPHASE is capturing the location-dependent genealogy of the data. Figure 5 *(bottom left, bottom right)* show the posterior distribution of the jump rate and the number of states, respectively. The spikes in the posterior of the jump rate are aligned with change points in the haplotype structure, indicating that recombination hotspots are accurately modelled by BNPPHASE.

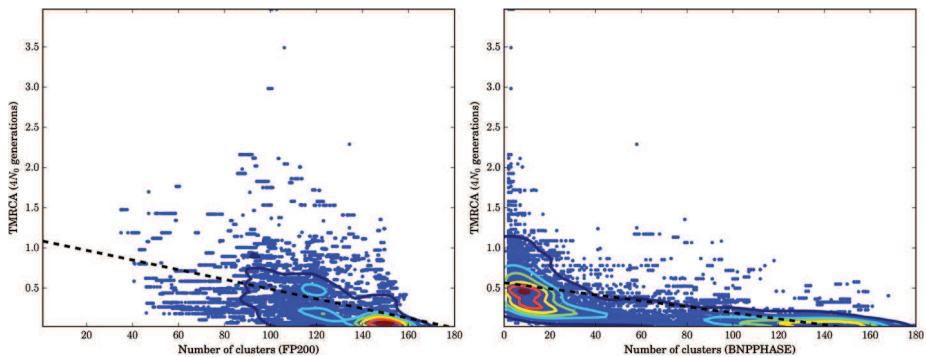

FIG 5. *Regression of TMRCA against number of clusters. Points indicate number of clusters at location (x-axis) vs TMRCA of location (y-axis). Contours show Gaussian kernel density estimate. Dotted line shows regression. Clusters found by* left: *fastPHASE with 200 clusters,* right: *BNPPHASE.*

Imputation results for the 20 regions of the male X chromosomes from the Thousand Genomes Project are shown in Figure 4. The BNPPHASE model consistently outperformed fastPHASE run with 10 components (the FP10 condition). For 30%, 50% and 90% hold out conditions, the performance of BNPPHASE and FP20 are quite similar. The BNPPHASE model tended to do better than other methods in larger hold out conditions. BEAGLE performed well on small hold out conditions, but poorly on large hold out conditions.

In a control designed to determine how important the priors placed on $\alpha, \alpha_0, \beta$ and $\gamma$ are, we considered two conditions for sampling the model parameters: a fixed condition in which the parameters were fixed to set values, and an unfixed condition in which hyperpriors were placed on the parameters. In the fixed condition, we perform a gridsearch over $\alpha, \alpha_0, \beta$ and $\gamma$ and we then ran MCMC chains with parameters fixed at the values that optimized test set accuracy for a fixed held out dataset. In the second condition (unfixed), MCMC was done for the full model, including slice sampling for $\alpha, \alpha_0, \beta$ and $\gamma$. The average accuracy of the fixed condition for the parameter values that maximized the gridsearch was 0.9917 whereas the average accuracy of the unfixed condition was 0.9919. Although small, this difference was found to be significant under a sign test ($p = 0.04$). All of the trends in Figure 4 were found to be significant under sign tests.

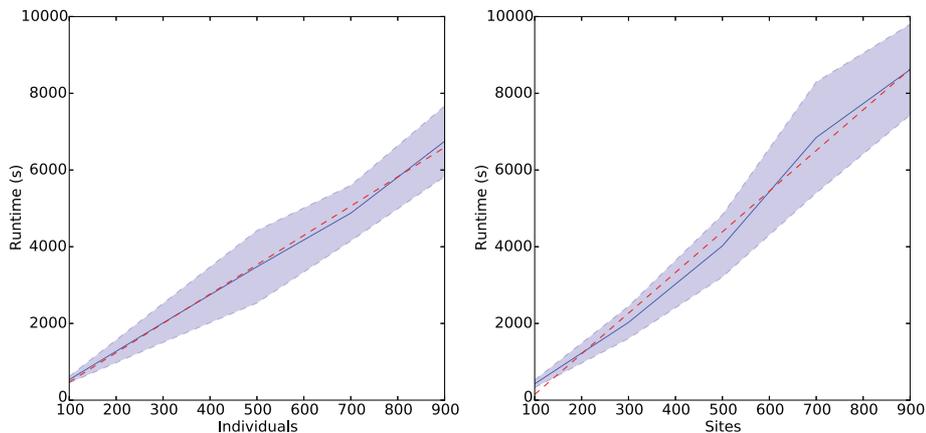

FIG 6. *Scalability of BNPPHASE. Runtime required for 200 iterations of BNPPHASE as* Left: *number of individuals or* Right: *number of sites is varied. Red dotted line indicates linear fit, solid blue line indicates mean over 10 trials, blue dotted line indicates standard deviation (shaded region is within one standard deviation of mean). Linear trend in both the number of individuals and the number of sites is clear.*

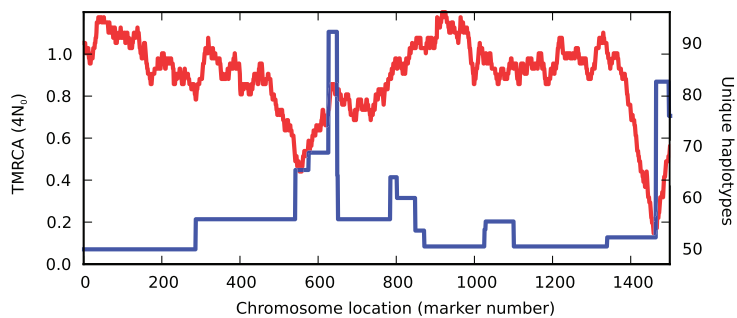

FIG 7. *Effect of TMRCA on number of haplotypes. Red plot indicates number of unique patterns per 100 location window. Blue plot indicates TMRCA.*

### 4.4. Experiment IIb) male X chromosomes, whole genome analysis

The accuracy of BNPPHASE and VBFP on the entire pseudo-autosomal region of the male X chromsomes from the Thousand Genomes Project is given in Table 2. The BNPPHASE model outperforms VBFP for all sizes of the VBFP parameter space that we considered. The difference in accuracies between BNPPHASE and VBFP20 is statistically significant under a Wilcoxon signed rank test (the $p = 0.045$). We note that the VBFP model with $K = 30$ performed similarly to the VBFP model with $K = 10$, but worse than the VBFP model with $K = 20$, which could indicate over fitting.

The median runtime of BNPPHASE on the 1251 X chromosome chunks was 37.52 minutes per chunk (the 25th and 75th runtime percentiles were 33.18 and 42.20 minutes, respectively). This was somewhat slower than the VBFP condition with 30 clusters, which had a median runtime of 22.45 minutes per chunk. This difference in runtime could be due to BNPPHASE using large numbers of clusters (more than 30) for some portions of the chromosome. In Figure 5 *(right)* we see that the for a small proportion of chromosome locations, the number of clusters used by the BNPPHASE model can be large.

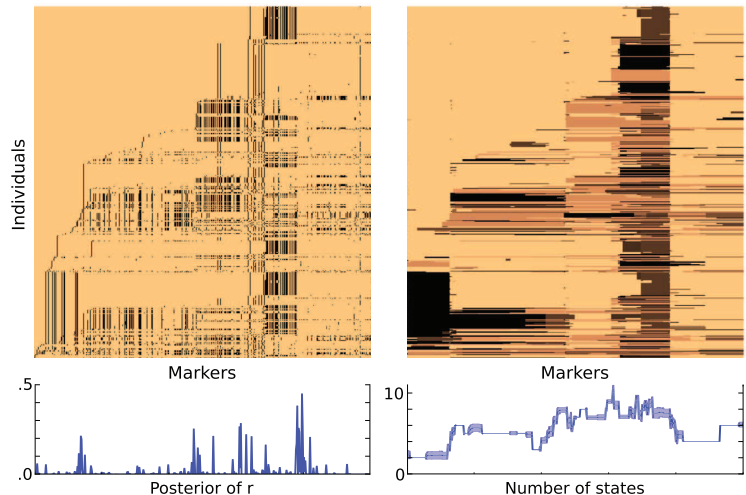

FIG 8. Top left: *Example region of male X chromosomes from the Thousand Genomes Project. x-axis indicates chromosome position, y-axis indicates individual identity. Black indicates minor alleles.* Top right: *Cluster assignment of sample from BNPPHASE model posterior. Color indicates cluster identity.* Bottom left, right: *Posterior distributions for r, and number of states averaged over MCMC samples (shaded region at sample standard deviation).*

TABLE 2
*Accuracy of BNPPHASE and variational Bayes version of fastPHASE on whole genome X chromosome analysis. Baseline accuracy found by calling major alleles at all locations. Mean proportion correct over 1251 chromosome chunks is reported, along with standard deviation in brackets.*

| Method | ACCURACY |
|---|---|
| BASELINE | 0.9517 (0.0202) |
| BNPPHASE | **0.9918** (0.0029) |
| VBFP10 | 0.9896 (0.0079) |
| VBFP20 | 0.9903 (0.0117) |
| VBFP30 | 0.9896 (0.0150) |

## 5. Discussion

We were surprised to see that the correlation between the number of clusters used by fastPHASE or BNPPHASE and the TMRCA was negative. This implies that there is more variation in recently coalesced material. In Figure 7, we explore whether or not this is reflected in the data by counting the num-ber of unique patterns of alleles in the simulated bottleneck from experiment *Ia)* in each 100 marker long window. We found that this empirical count was also significantly and negatively correlated with TMRCA (Pearson correlation $-0.7274$). This suggests that the regression coefficients found by BNPPHASE and fastPHASE are reasonable. Note that the size in basepairs of the windows used to determine the empirical counts around each location was variable (as the number of markers was fixed).

In experiment *IIb)*, there was some evidence that the parametric VBFP method was overfitting the data (moving from 20 to 30 clusters results in a drop of accuracy in out of sample prediction). Bayesian nonparametric priors can prevent overfitting by penalizing more complicated models in which the dimensionality of the latent space of the model is large. This could explain why the performance of BNPPHASE was better than VBFP on the whole genome analysis. The priors for $\theta$ and $r$ for VBFP and BNPPHASE were slightly different, but we do not expect that the VBFP result would change significantly if, for example, the heavy tailed prior for $r$ in BNPPHASE was to be included in the variational model.

## 6. Conclusion

We have provided a new Bayesian nonparametric model of genetic variation based on the hierarchical Dirichlet process. Unlike previous Bayesian nonparametric models, our model captures the nonhomogeneous structure of genetic variation in admixture and bottlenecks. We restricted the transitions of the HMM to support biologically plausible parameters, leading to more efficient inference and interpretability of model parameters. Further, the formulation of our model allows exact and truncation free Gibbs updates for the HMM clustering based on forwards filtering/backwards sampling. We showed that BNPPHASE provides imputation performance competitive with the state-of-the-art. Also, for simulated population bottleneck data, the BNPPHASE model provided more accurate regression against the TMRCA than the related fastPHASE model.

## Appendix: Message passing for $z_{it}, y_{it}$

Following from the notation and exposition in Section 2.4, we provide a message passing algorithm for an augmented Gibbs update for $z_{it}, y_{it}$. Note that $\zeta_{it} = 0 \Leftrightarrow y_{i,t-1} = 0$ and so an update for $z_{it}, y_{it}$ is fully described by an update for $z_{it}, \zeta_{it}$. In order to avoid instantiating all of the infinite vector $\omega$ (which, recall, is distributed according to the GEM distribution), we instantiate only the set $(\omega_k)_{k \in \mathcal{Z}^{-i}}$, where $\mathcal{Z}^{-i} = \{k : \exists t, i' \neq i : z_{i't} = k\}$ and we let $\omega_\varnothing = 1 - \sum_{k \in \mathcal{Z}^{-i}} \omega_k$ be the probability that $z$ joins a cluster that is not among any other assignments to coordinates of the GEM distribution. This allows us to represent the infinite object $\omega$ using a finite amount of memory, while still achieving exact updates.

By $\omega^{-i}$ we will refer to the components of $\omega$ that the blocks of $\mathcal{R}_t^{-i}$ are assigned to (*i.e.*, $\omega^{-i}$ is formed from $\omega$ by removing components that appear only among the component assignments of singleton blocks $\{i\}$ — blocks that were removed from $\mathcal{R}_t$ to form $\mathcal{R}_t^{-i}$). Since $\mathcal{R}_:^{-i}$, $\omega^{-i}$ and $\omega_\varnothing$ are determined by $(z_{i',:}, \zeta_{:,i'})$ for $i' \neq i$ and $\omega$, conditioning on 'rest' (the variables 'rest' are defined in Section 2.4) implies conditioning on $\mathcal{R}_:^{-i}$, $\omega^{-i}$ and $\omega_\varnothing$. (In this appendix, by indexing an object $A$ of length $T$ by : we refer to the vector $A_1, \ldots, A_T$.)

We will now consider the possible events that could occur when sequences $z_{i:}, \zeta_{i:}$ are sampled. We will augment the state space of $z_{i,:}, \zeta_{i,:}$ with the symbol $\varnothing$ and we will denote the event that sequence $i$ joins a singleton block at location $t$ by $\zeta_{it} = \varnothing$. In that case, the block will either be assigned to a component that already exists in $\omega^{-i}$, (which we will denote by $z_{it} = k$, where $k \geq 1$), or the block will be assigned to a new component (which we will denote by $z_{it} = \varnothing$). Conditioning on 'rest', the distribution of $z_{i,:}, \zeta_{i,:}$ is as follows:

$$\Pr(z_{i,:}, \zeta_{i,:}|\text{'rest'}) = J_1(z_{i1}, \zeta_{i1}) \prod_{t=2}^{T} \begin{cases} r_{t-1} J_t(z_{it}, \zeta_{it}) & \text{if } \zeta_{it} \neq 0, \varphi_{t\zeta_{it}} = z_{it}, \\ 1 - r_{t-1} & \text{if } \zeta_{it} = 0, z_{it} = z_{i,t-1}, \\ 0 & \text{otherwise.} \end{cases}$$

$$\cdot \prod_{t=1}^{T} L_{it}(x_{it}|z_{it}, \text{'rest'}). \quad (14)$$

$$\text{Where } J_t(z_{it}, \zeta_{it}) = \frac{1}{|R_t^{-i}| + \alpha} \cdot \begin{cases} |\zeta| & \zeta_{it} = \zeta \in \mathcal{R}_t^{-i}, \\ \alpha\omega_z, & \zeta_{it} = \varnothing, z_{it} = z \in \mathcal{Z}^{-i}, \\ \alpha\omega_\varnothing, & \zeta_{it} = z_{it} = \varnothing, \\ 0 & \text{otherwise.} \end{cases} \quad (15)$$

Here we have extended the definition $J_t(z, \zeta)$ from Section 2.4 to handle the augmentation of the state space of $z_{i,:}, \zeta_{i,:}$ with the symbol $\varnothing$. The definition of the likelihood $L_{it}$ is the same as that in Section 2.4, but with the following addition: if $z_{it} = \varnothing$, then $L_{it}(x_{it}|z_{it} = \varnothing, \text{'rest'}) = \beta_t^{x_{it}}(1 - \beta_t)^{1-x_{it}}$ (i.e., if a sequence joins a new cluster at location $t$, then there are no other sequences in that cluster at that location and so the likelihood of an allele is exactly given by the prior).

### 1. Gibbs update for cluster assignment of sequence $i$ in two steps

We will now provide a Gibbs sampling update for the $i$-th sequence based on the distribution (14) for $z_{i,:}, \zeta_{i,:}$ conditioned on the variables 'rest', in two steps. First, in Step 1 we will conduct forwards filtering/backwards sampling on $z_i, \zeta_i$ with the augmented state space for $z_{i,:}, \zeta_{i,:}$ described in the previous subsection. In the second step, for all $t$ with $z_{it} = \varnothing$ and $\zeta_{it} = \varnothing$, we assign clusters to $z_{it}$ and $\zeta_{it}$ through a retrospective stick breaking construction.

*Step 1: Forwards filtering/backwards sampling*

The forwards messages will be used in the forwards-filtering/backwards-sampling algorithm and the backwards messages will be used to compute marginal probabilities of an allele for imputation of missing data. Since $i$ is fixed, for the rest of the specification of Step 1 the subscript $i$ will be suppressed to make the notation more compact (so for example, by $z_t$ we will mean $z_{it}$). The messages are defined as follows:

$$\mathrm{m}_1^f(z_1, \zeta_1) = \Pr(x_1, z_1, \zeta_1 | \text{`rest'}),$$
$$\text{For } 1 < t \leq T: \ \mathrm{m}_t^f(z_t, \zeta_t) = \Pr(x_1 \ldots x_t, z_t, \zeta_t | \text{`rest'}),$$
$$\text{For } 1 \leq t < T: \ \mathrm{m}_t^b(z_t, \zeta_t) = \Pr(x_{t+1} \ldots x_T | z_t, \zeta_t, \text{`rest'}),$$
$$\mathrm{m}_T^b(z_T, \zeta_T) = 1. \qquad (16)$$

Due to the augmentation of $z_i, \zeta_i$ with $\varnothing$ and the condition $\zeta_{it} = 0 \Leftrightarrow y_{i,t-1} = 0$, for each of the messages in equations (16), the support of $(z_t, \zeta_t)$ is given by the following sets:

$$\text{For } t > 1: \ \sup(t) = \{(z, \zeta) : z = \varphi_{t\zeta}, \zeta \in \mathcal{R}_t^{-i}\}$$
$$\cup \{(z, \zeta) : z \in \{1, \ldots, K\}, \zeta = 0\}$$
$$\cup \{(z, \zeta) : z = \zeta = \varnothing\}.$$
$$\text{For } t = 1: \ \sup(1) = \{(z, \zeta) : z = \varphi_{1\zeta}, \zeta \in \mathcal{R}_1^{-i}\}$$
$$\cup \{(z, \zeta) : z = \zeta = \varnothing\}. \qquad (17)$$

For the backwards messages, if we condition on $z_t$ then $\zeta_t$ and $x_{t+1}, \ldots, x_T$ are independent, and so for a fixed $z$ the messages $\mathrm{m}_t^b(z, \zeta)$ all have the same value for each $\zeta : (z, \zeta) \in \sup(t)$ and therefore we will refer to this value by $\mathrm{m}_t^b(z)$. Further, we will often find it useful to sum the forwards messages over the possible values of their parameters and so we will introduce the following shorthand notations:

$$\mathrm{m}_t^f = \sum_{(z,\zeta) \in \sup(t)} \mathrm{m}_t^f(z, \zeta).$$

And for a fixed $z$ with $0 \leq z \leq K$:

$$\mathrm{m}_t^f(z) = \sum_{\substack{(z',\zeta) \in \sup(t), \\ z'=z}} \mathrm{m}_t^f(z', \zeta).$$

The forwards messages in display (16) can be computed recursively as follows:

$$\mathrm{m}_1^f(z_1, \zeta_1) = L_1(x_1 | z_1, \text{`rest'}) J_1(z_1, \zeta_1).$$

And for $1 < t \leq T$:

$$\mathrm{m}_t^f(z_t, \zeta_t) = L_t(x_t | z_t) \cdot \begin{cases} (1 - r_{t-1}) \cdot \mathrm{m}_{t-1}^f(z_t) & \text{if } \zeta_t = 0, \\ r_{t-1} \cdot \mathrm{m}_{t-1}^f \cdot J_t(z_t, \zeta_t) & \text{otherwise.} \end{cases} \qquad (18)$$

Here $J_t$ is the same function defined previously in this appendix. In a similar way, the backwards messages can also be computed recursively as follows:

$$\mathrm{m}_t^b(z_t) = (1 - r_t) L_t(x_{t+1} | z_t) \mathrm{m}_{t+1}^b(z_t) + r_t \sum_{(z,\zeta) \in \sup(t+1)} L_{t+1}(x_{t+1} | z) J_{t+1}(z, \zeta) \mathrm{m}_{t+1}^b(z).$$
$$(19)$$

After computing the forwards messages, the cluster assignments for the $i$-th sequence can be sampled through a backwards-sampling algorithm. By Bayes' rule, the Markov property of the cluster assignments and the definition of the forwards messages, the probabilities for this backwards sampling are as follows:

$$\Pr(x, z_T, \zeta_T|\text{'rest'}) \propto \mathrm{m}_T^f(z_T, \zeta_T)$$
$$\Pr(x, z_t, \zeta_t|z_{t+1}, \zeta_{t+1}, \text{'rest'}) \propto \Pr(x_1, \ldots, x_t, z_t, \zeta_t|\text{'rest'})$$
$$\cdot \Pr(z_{t+1}, \zeta_{t+1}|z_t, \zeta_t, \text{'rest'}),$$
$$\propto \mathrm{m}_t^f(z_t, \zeta_t) \cdot \begin{cases} \delta(z_t = z_{t+1}) & \text{if } \zeta_{t+1} = 0, \\ 1 & \text{otherwise.} \end{cases} \quad (20)$$

Here $\delta$ denotes the Kronecker delta function, so $\delta(z_t = z_{t+1})$ is one if $z_t = z_{t+1}$ and zero otherwise. In equation (20), the domain of $z_t, \zeta_t$ is always restricted to $\sup(t)$. Step 1 of the Gibbs update for $z_i, \zeta_i$ is thus given by sampling $z_t, \zeta_t$, recursively in descending order of $t$ using the probabilities given in display (20).

While the backwards messages are not required for sampling equation (20), in order to predict held out or unobserved alleles, the backwards messages are computed and combined with the fowards messages in order to find allele probabilities, with the cluster assignment marginalized. This is standard in MCMC methods for HMM inference.

*Step 2: Retrospective stick breaking*

We now provide a retrospective stick breaking scheme to select the components for the singleton blocks which were sampled in Step 1 but whose assigned components were not in $\omega^{-i}$. That is, we will now sample the values $z_{it}$ for $t$ such that, after Step 1, $z_{it} = \varnothing$. We will refer to such $t$ by the set $S_\varnothing = \{t : z_{it} = \varnothing\}$. For a given setting of $z_{i,:}, \zeta_{i,:}$ sampled using the backwards-sampling from Step 1, $S_\varnothing$ is found deterministically. Applying Step 1 followed by Step 2 yields a full Gibbs update for $z_{i,:}, \zeta_{i,:}$.

By the definition of the symbol $\varnothing$, the variables $z_{it} : t \in S_\varnothing$ should only be assigned to components of the GEM $\omega$ that none of the other block in $\mathcal{R}_:^{-i}$ are assigned to. It is, however, possible for more than one $z_{it} : t \in S_\varnothing$ to be assigned to the same component. For each $t \in S_\varnothing$, $z_{it}$ marginally follows the law $\Pr(z_{it}|\text{'rest'}, z_{it} \notin \omega^{-i})$. Since $i$ is fixed, for a fixed location $t$ there is at most one $z_{it}$ that needs to be sampled for $t$, and so the allele counts $n_{1tk}^{-i}, n_{0tk}^{-i}$ are conditionally independent of the random variables $z_{it'} : t' \neq t$ (conditioned on 'rest'). Further, because $\zeta_{it} \neq 0$ for all $t \in S_\varnothing$, the $z_{it} : t \in S_\varnothing$ are independent conditioned on the GEM $\omega$. Combining these two observations, it is clear that $z_{it} : t \in S_\varnothing$ is sampled i.i.d. directly from the prior of the BNPPHASE model conditioned on the event that $z_{it} \notin \mathcal{Z}^{-i}$. This can be done by using the stick breaking construction in equation (2) to instantiate components of $\omega$ that are not in $\omega^{-i}$ (we will refer to these components by $\omega_{\varnothing 1}, \omega_{\varnothing 2}, \ldots$) and then sampling $z_{it} : t \in S_\varnothing$ from $\omega$ restricted to these new components. This can be done efficiently by sampling a uniform variate $u_t$ i.i.d. for each $t \in S_\varnothing$ and then setting $z_{it}$ to the smallest $k$ such that:

$$\frac{\sum_{k'=1}^{k} \omega_{\varnothing k'}}{\omega_{\varnothing}} > u_t \tag{21}$$

With this scheme, $\omega_{\varnothing 1}, \omega_{\varnothing 2}, \ldots$ can be sampled in sequence, stopping as soon as equation (21) is satisfied for all $t \in S_{\varnothing}$. Step 2 is made explicit in the following enumeration, which should be performed immediately after sampling $z_{i,:}, \zeta_{i,:}$ according to Step 1.

1. Set $S_{\varnothing} \leftarrow \{t : z_{it} = \varnothing\}$.
2. Set $\omega'_{\varnothing,:} \leftarrow ()$ and set $K'_{\varnothing} \leftarrow 0$.
3. For each $t \in S_{\varnothing}$:
    (a) Set $\mathcal{R}_t \leftarrow \mathcal{R}_t^{-i} \cup \{\{i\}\}$ and set $\zeta_{it} \leftarrow \{i\}$.
    (b) Draw $u \sim \text{Uniform}(0, 1)$.
    (c) While $k^* = \min\{1 \leq k \leq K'_{\varnothing} : \sum_{k'=1}^{k} \omega_{\varnothing k'} > u\}$ does not exist:
        i. Set $K'_{\varnothing} \leftarrow K_{\varnothing} + 1$.
        ii. Draw $\nu \sim \text{Beta}(1, \alpha)$.
        iii. Set $\omega'_{\varnothing, K'_{\varnothing}} \leftarrow \nu \left(1 - \sum_{k=1}^{K'_{\varnothing}-1} \omega'_{\varnothing k}\right)$.
    (d) Set $z_{it} \leftarrow k^*$.
4. Set $\omega_{\varnothing,:} \leftarrow ()$ and set $K_{\varnothing} \leftarrow 0$.
5. For $k' = 1$ to $K'_{\varnothing}$:
    (a) If there exists $t \in S_{\varnothing}$ such that $z_{it} = k'$:
    (b) Set $S \leftarrow \{t \in S_{\varnothing} : z_{it} = k'\}$.
    (c) If $|S| > 0$:
        i. Set $K_{\varnothing} \leftarrow K_{\varnothing} + 1$.
        ii. Set $z_{it} \leftarrow K^{-i} + K_{\varnothing}$ for each $t \in S$.
        iii. Set $\omega_{\varnothing, K_{\varnothing}} \leftarrow \omega_{\varnothing} \cdot \omega'_{\varnothing, k'}$.
        iv. Set $S_{\varnothing} \leftarrow S_{\varnothing} \setminus S$.
6. Set $\omega \leftarrow ((\omega_k^{-i})_{k=1}^{K^{-i}}, (\omega_{\varnothing, k})_{k=1}^{K_{\varnothing}})$ and set $K \leftarrow K^{-i} + K_{\varnothing}$.
7. Set $K \leftarrow K + K_{\varnothing}$
8. Set $\omega_{\varnothing} \leftarrow 1 - \sum_{k=1}^{K} \omega_k$.

The concatenation of Step 1 and Step 2 provides a full Gibbs update for the latent cluster assignment of the $i$-th sequence.

### *Acknowledgements*

We thank the Gatsby Charitable Foundation for funding. We also thank Chris Holmes and Ioanna Manolopoulou for helpful discussion.